\documentclass[pdflatex,sn-mathphys-num]{sn-jnl}

\usepackage{graphicx}%
\usepackage{multirow}%
\usepackage{amsmath,amssymb,amsfonts}%
\usepackage{amsthm}%
\usepackage{mathrsfs}%
\usepackage[title]{appendix}%
\usepackage{xcolor}%
\usepackage{textcomp}%
\usepackage{manyfoot}%
\usepackage{booktabs}%
\usepackage{algorithm}%
\usepackage{algorithmicx}%
\usepackage{algpseudocode}%
\usepackage{listings}%

\usepackage{natbib}

\usepackage{xspace}

\theoremstyle{thmstyleone}%
\theoremstyle{thmstyletwo}%

\theoremstyle{thmstylethree}%

\def\Fref#1{Figure~\ref{#1}\xspace}

\begin{document}

\title[Strong gravitational lensing with Euclid]{Strong gravitational lensing with Euclid}


\author*[1]{\fnm{Alessandro} \sur{Sonnenfeld}}\email{sonnenfeld@sjtu.edu.cn}

\affil*[1]{\orgdiv{Department of Astronomy, School of Physics and Astronomy}, \orgname{Shanghai Jiao Tong University}, \orgaddress{\street{800 Dongchuan Road}, \city{Shanghai}, \postcode{200240}, \country{China}}}


\abstract{
Euclid data is going to lead to the discovery of approximately $10^5$ new strong gravitational lenses, thereby increasing the number of known lenses by a factor of $30$.
Such an impressive jump has the potential to revolutionise the field of strong lensing, by covering an ever-increasing range in the parameter space occupied by the lenses and lensed sources, and by uncovering rare and precious strong lens systems.
The first quick data release of Euclid has already revealed hundreds of lenses, giving us a taste of what the final Euclid strong lens sample might look like.
These lenses include a variety of objects acting as deflectors, from isolated galaxies to clusters, from early-type galaxies to edge-on disks. As many as four double-source plane lenses, in which one foreground object lenses two galaxies at two different redshifts, have already been discovered.
As more data are acquired, it might become possible to detect the strong lensing signature of dwarf galaxy satellites, and to find lenses with an unusually large number of images.
In this paper, we try to anticipate the kind of discoveries that these data will bring along.
}




\maketitle

\section{Introduction}\label{sect:intro}

Strong gravitational lensing, the production of multiple images of individual light sources lensed by massive objects, is one of the most fascinating phenomena observed in the sky.
The number of known strong lenses has been growing steadily since the discovery of the first multiply-imaged quasar in 1979 \cite{WCW79}, and is now up in the thousands.
The lenses discovered thus far sample a broad range in the properties of the (foreground) lens objects and of the (background) lensed sources.
Lenses with well-resolved systems of multiple images have masses that span approximately five orders of magnitude, from $\sim10^{15}M_\odot$ of the most massive clusters of galaxies \cite{Kne++93}, to $\sim10^{10}M_\odot$ of dwarf galaxy satellites \cite{VCK10}. 
In between, massive early-type galaxies make up the bulk of the population of deflectors, but tens of disk galaxy lenses have been discovered as well \cite{Tre++11}.
While the majority of the background sources are optically-bright star-forming galaxies, known lensed objects include quasars, quiescent galaxies \cite{Aug++11}, dusty galaxies detected via molecular line emission \cite{Hez++13}, and supernovae \cite{Kel++15}.

A major turning point in strong lensing research has been the advent of the Hubble Space Telescope (HST). High-resolution space-based images from HST have been essential for the discovery and study of hundreds of strong lenses, both galaxy- and cluster-scale ones.
Highlights from strong lensing research based on HST imaging data include: 
the first lensing-based systematic studies of the internal structure of early-type galaxies \cite{Koo++06,Aug++10};
the discovery of the first double-source plane lens, with two sources at two different redshifts lensed into two concentric Einstein rings \cite{Gav++08};
the detection of a mass clump with no luminous counterpart, close in projection to a strong lens \cite{Veg++10b}; 
measurements of the expansion rate of the Universe from the time delay between multiple images of lensed variable sources \cite{Suy++10,Kel++23,Gri++24};
measurements of the dark matter distribution in the inner regions of galaxy clusters \cite{New++13};
and constraints on the galaxy luminosity function up to $z\sim10$, obtained by exploiting the magnifying power of galaxy clusters \cite{Coe++13,Ish++18}. 
The James Webb Space Telescope (JWST) is currently enabling further progress, especially in the study of lensed sources \cite{Rob++23,Mow++24}. 

Yet, the relatively small field of view of HST and JWST instruments sets a hard limit to the type of investigations that can be carried out with these observatories.
HST and JWST will only ever observe a small fraction of the full sky over their lifetime. 
As strong lensing is a rare phenomenon, this makes it difficult to obtain high statistics over the entire parameter space of interesting lens properties, and, without a high-resolution survey of a large part of the sky, some of the most peculiar systems might be missed altogether.

The Euclid satellite, launched on July 1st, 2023, is providing us with such a survey.
Euclid has a primary mirror of $1.2$ metres in diameter, and is equipped with two instruments for its scientific studies, VIS and NISP. VIS is a single-filter optical camera, with a field of view of $0.57$ square degrees, and $0.1''$ pixels.
Its fine point spread function (PSF), with a $0.13''$ FWHM, allows for the clean identification of strong lenses with image separation at least down to $1''$ \cite{SLDEA}.
NISP is a near-infrared spectrograph and camera. Although its images are not as sharp as those obtained by VIS (its pixels are $0.3''$ in size), it can provide valuable information for the discovery and analysis of strong lenses.
Euclid is expected to observe $14000$ square degrees, one third of the full sky, down to an AB magnitude of $24.5$ at optical wavelengths, over the course of five years \cite{Mel++25}.
At the time of writing of this paper, results from the first quick data release \cite[Q1;][]{Q1}, which covers an area of $63$ square degrees, have been published. 

Approximately $600$ new galaxy-scale lenses and $83$ cluster-scale ones have so far been discovered in Q1 \cite{SLDEA,SLDEB,SLDEC,SLDED,SLDEE,SLDEF,Xu++26,Ber++26}, making strong lensing one of the highlights of the early science activities of Euclid. 
Extrapolating this rate to the full Euclid survey, we expect that approximately $10^5$ lenses will eventually be found in its data. 
In the next section, we will try to anticipate what kind of discoveries can be made with these lenses.
While far from complete, this list focuses on areas in which Euclid observations can provide a distinct advantage over surveys from other facilities.

\section{Outlook after the Q1 data}\label{sect:stats}

\subsection{A new window on galaxy evolution}

\begin{figure}
\includegraphics[width=\textwidth]{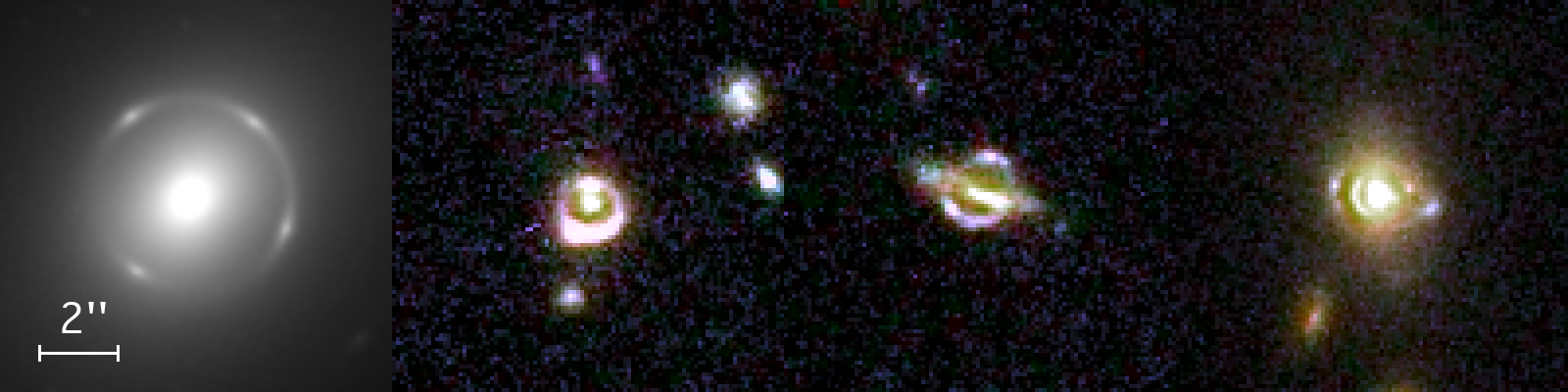}
\caption{
Examples of strong lenses discovered in Euclid data so far. 
From left to right: NGC6505, EUCL\,J040234.43$-$500554.2, EUCL\,J040823.49$-$500956.2, and EUCL\,J181422.71$+$670816.4, also known as the `teapot lens' \cite{Ori++25,SLDEA,SLDED}.
For NGC6505 we are showing a VIS-only greyscale image, while for the other lenses, VIS and NISP colour-composite images are shown.
The redshift of NGC6505 is $z=0.042$, while EUCL\,J040234.43$-$500554.2 has a lens photo-$z$ of $\sim~1.1$. 
\label{fig:examples}
}
\end{figure}

Most of our current understanding of galaxy evolution relies on observations of the light-emitting component of galaxies, stars and gas.
With strong lensing, we have the opportunity to add a new dimension to the investigation: mass.
The lenses discovered in Euclid data span already a remarkably broad redshift range: from NGC~6505 at $z=0.042$ \cite{Ori++25} to lens galaxies with photo-$z$s of $\sim~1.1$ (see \Fref{fig:examples}). 
With a skilful use of the NISP data, lenses at redshifts of $2$ and beyond could soon be found.
The main observable quantity of a strong lens is its Einstein radius, which is roughly the radius of the circle that encompasses the multiple images of the source.
A single lens can provide a very precise estimate of the total mass enclosed within the Einstein radius, with a typical uncertainty of a few percent, provided that the redshift of the lens galaxy and background source are known.
The statistical combination of a large set of such measurements can tell us how the internal structure of massive quiescent galaxies, which make up the bulk of the existing lens sample, has evolved in the last 10 billion years.
We know, from observations of the stellar component, that the average size of quiescent galaxies increases with time \citep{vDo++10}. In other words, galaxies are more extended today than they were at redshift $2$.
We expect the mass distribution to follow the same trend: galaxies at low redshift should have a shallower density profile than their higher redshift counterparts, as a result of dissipationless (dry) mergers \cite{SNT14,Rem++17,Wan++19}. So far, strong lensing measurements favour the opposite trend \cite{Bol++12,Son++13}, though that might be the result of selection effects \cite{Son24}.
Euclid data give us the opportunity to extend the investigation of the density profile to galaxies in the range $1<z<2$, where substantial evolution of the massive quiescent population should still be taking place \cite{New++12}. 
An agreement between Euclid measurements and theoretical predictions would signify an important test passed by current models for galaxy growth. Conversely, if the discrepancy that emerged from early strong lensing measurements were to be confirmed by Euclid, a rethinking of the dry merger-dominated late evolution of galaxies will be in order.

A more ambitious goal would be to separately constrain the stellar and dark matter mass in massive galaxies. It might be possible to do so by combining strong lensing observations with weak lensing constraints on larger scales \citep{ZSH24}. Accurate stellar mass measurements are direly needed, as current ones are plagued by the systematic uncertainty associated with the choice of the stellar initial mass function (IMF). 
Our knowledge of the dark matter distribution is even more uncertain, as current measurements of it are scarce.
Constraints from stellar dynamics, in isolation \cite{Cap++13} or in combination with strong lensing \cite{Sha++21}, seem to point at lower dark matter fractions compared to hydrodynamical simulations \cite{Sha++21}. If confirmed by Euclid, this would require a thorough revision of the subgrid physical models used in simulations, or even a questioning of the collisionless cold dark matter (CDM) paradigm. 
Nevertheless, given our current ignorance of the dark matter distribution in galaxies, a lensing-based measurement of it from Euclid has ample potential for a disruptive discovery.

\subsection{Disk lenses: the next chapter in the dark matter vs. MOND debate?}

While the primary observable is the Einstein radius, strong lensing data are also sensitive to the azimuthal mass structure of a lens. 
This feature is particularly useful for edge-on disk lenses, because their two dominant mass components, the disk and the dark matter halo, have radically different imprints on the overall flattening of the total projected mass.
In other words, it is possible via lens modelling to disentangle the contribution of the disk and the dark matter halo to the total mass. 
Roughly $5\%$ of the Euclid lenses discovered so far have an edge-on disk as the deflector (see \Fref{fig:examples} for an example), which extrapolate to a few thousand lenses over the full area. 
Recent measurements based on rotation curve data seem to indicate that disk galaxies have a lower than expected dark matter fraction \cite{Pos++19}.
A statistical analysis of a large sample of disk lenses from Euclid can provide a completely independent view of the problem. 
Moreover, strong lensing measurements can reveal if the dark matter halo is flattened, as predicted by models based on the collisionless CDM paradigm \cite{A+B10}, and to what extent. 

Disk lenses also make for excellent testbeds for alternative models of gravity with no dark matter, such as Modified Newtonian Dynamics (MOND).
This is because mass-follows-light models and models with a spheroidal dark matter halo predict very different strong lensing signals around disk galaxies \cite{H+G25}, as a consequence of the difference in the azimuthal distributions of dark matter and light.
So far, the most heated debates between dark matter and MOND proponents have revolved around merging clusters of galaxies, but observations from such systems can be difficult to interpret due to the complex astrophysical processes occurring within them. Edge-on disk lenses can provide a cleaner alternative.
It is therefore likely that the next chapter in the MOND debate will be based on disk strong lenses.

\subsection{The inner structure of satellite galaxies in clusters}

Clusters of galaxies make for some of the most spectacular manifestations of the strong lensing phenomenon.
Typically, the multiple images of sources lensed by clusters have separations well into the tens of arcseconds, enclosing dozens of cluster member galaxies within them.
Occasionally, however, individual satellite galaxies located outside of the main critical curve can produce their own Einstein ring \citep[see e.g.][]{Par++16,Den++25}.
These galaxy-scale lenses can be used to measure the inner structure of satellite galaxies, and to get a better understanding of dynamical processes that are important in galaxy clusters, such as tidal stripping.
Recently, \cite{Men++20} claimed, on the basis of the detailed model of two strong lensing clusters, that cluster satellites account for a significantly larger strong lensing cross-section than predicted by simulations for similar objects.
In other words, the rate of galaxy-scale lensing events in observed clusters appears to be larger than in simulated ones.
This observation has so far proven to be difficult to reconcile with the $\Lambda$CDM paradigm \cite{Tok++24}.
However, current constraints on the lensing cross-section of satellite galaxies are indirect: they are based on the perturbations induced on the image positions of sources lensed by the whole cluster, and not on the actual number of galaxy-scale strong lensing events, that are too few to draw strong conclusions.
Euclid will detect thousands of galaxy-scale strong lenses in clusters. 
A statistical analysis of those lenses will provide a direct measurement of the lensing cross-section of satellite galaxies, and a much more solid measurement of their inner structure.
If the discrepancy with simulations will persist, it could constitute a challenge for the $\Lambda$CDM model.

\subsection{Satellite dwarf galaxies and their masses}

\begin{figure}
\begin{tabular}{cc}
\includegraphics[width=0.45\textwidth]{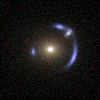} & 
\includegraphics[width=0.45\textwidth]{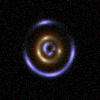}
\end{tabular}
\caption{
Example simulated lenses.
Left: galaxy-scale lens with a satellite perturber. The satellite produces its own miniature Einstein ring, which causes the main arc to split.
Right: zig-zag double-source plane lens. The outermost and innermost blue rings are multiple images of the same background galaxy, strongly lensed by two galaxies. The middle white ring is the second of the lenses, as counted from the observer. Light from the foreground lens is not shown, to highlight lensed features.
For the sake of illustration, these colour-composite images have the same PSF in each channel (in contrast to Euclid, where pixels of the NISP detector are three times the size of the VIS ones).
\label{fig:sims}
}
\end{figure}

Similarly to clusters, the substructure associated with individual galaxies can also be probed with Euclid strong lensing.
If a satellite happens to lie in projected proximity to the main galaxy acting as a lens, its gravitational field can produce a visible perturbation on the surface brightness distribution of a strongly lensed image.
The left-hand panel of \Fref{fig:sims} shows a simulated example of this effect \citep[see][for real-world cases]{Veg++10a,S+H10}.
For a satellite to produce an Einstein ring that can be resolved by Euclid, its mass must be on the order of $10^{10}M_\odot$ or larger.
If located at cosmological distances, such galaxies are very difficult to study in any way other than strong lensing.
Euclid observations can therefore prove to be extremely valuable for the study of the inner structure of dwarf galaxies and its evolution.

\subsection{Science with multiple source plane lenses}

A typical strong lens consists of a pair of galaxies, a lens and a source.
But, in rare cases, three or more galaxies can be found near the same line of sight, producing lenses that act on multiple source planes.
Four double-source plane lenses have already been discovered in Euclid Q1 \cite{SLDED}, meaning that hundreds of them are expected to be found by the end of the survey.
Double-source plane lenses can provide more stringent constraints on the mass distribution of the lens galaxy at the front, compared to regular lenses.
They could be used to obtain precise measurements of the stellar and dark matter mass in individual lens galaxies. This would be particularly useful for constraining the scatter in these properties, which is notoriously difficult to measure, and which would complement well the information coming from the statistical analysis of large samples of regular lenses.
Double-source plane lenses are also sensitive to cosmology, as the ratio between the the Einstein radii of the two sources scales with a ratio of angular diameter distances between the three galaxies, at fixed lens density profile \citep{Gav++08}. However, changes in cosmology can be absorbed by corresponding adjustments to the mass distribution of the two lenses \citep{Sch14}. This means that additional data other than strong lensing are likely required to break this degeneracy.

Euclid might uncover even rarer and more valuable types of lenses.
One such type is an Einstein zig-zag, which consists of a double-source plane lens in which light rays from the furthest source travel around opposite sides of the two lenses, producing additional images near the centre \citep[see][for an illustration of the geometry of the system]{WAE08}. The right-hand panel of \Fref{fig:sims} shows a simulated example of an Einstein zig-zag lens (light from the lens at the front is not shown, to highlight the ring structure).
A lens of this kind provides even more constraints on the mass density profile of the two lenses, potentially allowing for some of the most detailed determinations of the inner structure of a galaxy.
Crucially, the innermost ring in an Einstein zig-zag is not as de-magnified as central images in regular lenses typically are, and therefore could be used to measure the mass of the central supermassive black hole in both lenses. 
The example shown in \Fref{fig:sims} is only one of many types of image configurations that double source plane lenses can produce.
We refer to \cite{C+B16} for a compendium of such possible images, along with their relative probabilities of occurring.

In order to produce a zig-zag image, the two lenses must be of comparable mass. For this reason, we expect only a small fraction of double source plane lenses to be Einstein zig-zags. 
Nevertheless, one zig-zag lens is already known, in the form of a six-image strongly lensed quasar \citep{Dux++25,Sch++25}.
Therefore, it is likely that more lenses of a similar kind will be discovered in Euclid data.


\section{Conclusions}\label{sect:concl}

The above is a partial list of the kind of strong lensing data that Euclid will produce, and the discoveries that those data will enable. 
Euclid will also deliver images of giant arcs, lensed quasars, and lensed high-redshift sources, allowing for a broader variety of science investigations than can be properly addressed in this paper.
Euclid data can also play a crucial role in providing lens models for the analysis of lensed transient events, such as supernovae, found with other facilities.

One challenge that the strong lensing community will be facing in the exploitation of the high-resolution images of strong lenses from Euclid is the availability of spectroscopic data.
Spectroscopy has been one of the key factors in determining the scientific impact of past strong lensing campaigns. For instance, the success of the Sloan ACS Lens Survey \citep[SLACS;][]{Bol++06,Tre++06,Koo++06}, owes in large part to the fact that it was designed as a spectroscopy-first survey.
By construction, each lens in the SLACS sample has a spectroscopic redshift of the lens and the source galaxy, allowing for the prompt conversion of angular quantities, such as the Einstein radius, into precise measurements of mass.
With Euclid, spectroscopic information will realistically be available for only a small fraction of the full strong lens sample. Euclid carries a slitless spectrograph on the NISP instrument, but it is mostly sensitive to emission lines from the brightest sources over a limited redshift range. The 4MOST spectroscopic survey \citep{deJ++19} is currently carrying out a follow-up campaign of $10\,000$ Euclid lenses \citep{Col++23}, but it focuses on a sub-region of the flux--redshift distribution of the lenses and lensed sources. 
In practice, this means that, in order to carry out the investigations described in this paper, additional spectroscopic observations will be needed.
The impact of Euclid on the field of strong lensing will therefore depend, in part, on the willingness of the astronomical community to allocate telescope time to the follow-up strong lens targets, as well as on the ability of the strong lensing community to prioritise spectroscopic observations in an optimal way, and make the most of the available data.

Nevertheless, Euclid will open up numerous new opportunities in the field of strong gravitational lensing. 
The catalogue of Euclid strong lenses is set to become the main reference for strong lensing experts in years to come. 
We hope that the astronomical community will find ingenious ways of using these data, and contribute to the range of surprises that will come with them.

\backmatter

\bmhead{Acknowledgements}

This work has made use of the Euclid Q1 data from the {\it Euclid} mission of the European Space Agency (ESA), 2025,
\url{[https://doi.org/10.57780/esa-2853f3b|https://doi.org/10.57780/esa-2853f3b]}, and ESA Datalabs (\url{datalabs.esa.int}), an initiative by ESA's Data Science and Archives Division in the Science and Operations Department, Directorate of Science.
The author acknowledges a sponsorship from the Yangyang Development Fund, and from the National Key R\&D Program of China (No. 2023YFA1607800, 2023YFA1607802).

\bibliography{truncated_references}


\end{document}